\documentclass[a4paper,11pt,twoside]{article}  

\usepackage{graphics,epsfig,pstricks}

\usepackage{amsmath,amsbsy}
\usepackage{amssymb}
\usepackage{mathrsfs}

\usepackage{natbib}

\newcommand{\nompropre}[1]{#1}
\newcommand{\oldstyle}[1]{#1}

\newcommand{\imat}{\mathrm{i}}
\newcommand{\dmat}{\mathrm{d}}

\newcommand{\RR}{\mathbb{R}}
\newcommand{\ZZ}{\mathbb{Z}}

\newcommand{\EXP}[1]{\mathrm{e}^{#1}} 

\newcommand{\tv}[1]{\boldsymbol{#1}}

\newcommand{\DEF}{\overset{\mathrm{def}}{=}}
\newcommand{\DEFt}{\smash{\overset{\text{\tiny def}}{=}}}

\newcommand{\Deg}{\text{\textsc{l}}}
 \newcommand{\Dim}{\text{\textsc{d}}}

\newcommand{\at}{{\char '100}}

\newcommand{\ket}[1]{|#1\rangle} 
\newcommand{\kete}[1]{|\kern.3ex#1\kern.3ex\rangle}
\newcommand{\bra}[1]{\langle #1 |}
\newcommand{\brae}[1]{\langle\kern.3ex #1 \kern.3ex|} 
\def\braket#1#2{\mathinner{\langle{#1}|{#2}\rangle}}

\makeatletter
\newsavebox\myboxA
\newsavebox\myboxB
\newlength\mylenA

\newcommand*\xoverline[2][0.75]{%
    \sbox{\myboxA}{$\m@th#2$}%
    \setbox\myboxB\null
    \ht\myboxB=\ht\myboxA%
    \dp\myboxB=\dp\myboxA%
    \wd\myboxB=#1\wd\myboxA
    \sbox\myboxB{$\m@th\overline{\copy\myboxB}$}
    \setlength\mylenA{\the\wd\myboxA}
    \addtolength\mylenA{-\the\wd\myboxB}%
    \ifdim\wd\myboxB<\wd\myboxA%
       \rlap{\hskip 0.5\mylenA\usebox\myboxB}{\usebox\myboxA}%
    \else
        \hskip -0.5\mylenA\rlap{\usebox\myboxA}{\hskip 0.5\mylenA\usebox\myboxB}%
    \fi}
\makeatother

\newcommand{\barQ}{\bar{Q}}
\newcommand{\barK}{\bar{K}}
\newcommand{\barU}{\bar{U}}

\newcommand{\hatW}{\,\widehat{\!W\!}\,}

\newcommand{\cc}{\cdot}

\begin{document}

\title{Path integrals in a multiply-connected configuration space (50 years after)}
\author{Amaury Mouchet}

\date{Institut Denis Poisson de Math\'ematiques
  et de  Physique Th\'eorique, Universit\'e de Tours --- \textsc{\textsc{cnrs (umr 7013)}}\\
 Parc de Grandmont 37200
 Tours,  France.\\ {mouchet\at phys.univ-tours.fr}\\
Version 2.0}


\maketitle

\begin{abstract}
The proposal made 50 years ago by \citet{Schulman68a},
\citet{Laidlaw/MoretteDeWitt71a} and \citet{Dowker72a} to
decompose the propagator according to the homotopy classes of paths
was a major breakthrough: it showed how Feynman functional integrals
opened a direct window on quantum properties of topological origin in
the configuration space.  This paper casts a critical look at the
arguments brought by this series of papers and its numerous followers
in an attempt to clarify the reason why the emergence of the unitary
linear representation of the first homotopy group is not only
sufficient but also necessary.
\end{abstract}

\begin{flushright}
 \textit{\small We must neglect our models and study our capabilities.}\\\hfill\small
\nompropre{Edgar Allan} \citet[p.~122]{Poe1845a}.
\end{flushright}

This article comes back to the 50-years-old following statement: when
the quantum propagator in configuration space is split into homotopy
classes of paths according to
\begin{equation}\label{eq:Kdecomp}
  K(q_f,t_f,q_i,t_i)=\sum_{\mathfrak{c}\in\pi_1(q_i,q_f)}E(\mathfrak{c})\int_{\mathscr{C}\in\mathfrak{c}}
\EXP{\frac{\imat}{\hbar}S[\mathscr{C}]}\dmat[\mathscr{C}]\;,
\end{equation} 
then the coefficients~$E(\mathfrak{c})$ are \emph{necessarily} given by the
images of a unitary representation of the first homotopy group of the
configuration space.

After having presented the context and the high stakes of such
decomposition, section~\ref{sec:intro} will briefly recall the basic concepts while setting up the
notations. The core of this work will be the object of section~\ref{sec:Kdecomp}
where a proof of this statement will be proposed. This will provide us
a vantage point from which we will be able to cast, in section~\ref{sec:SLdWB}, a
critical eye on the arguments advocated by \citet{Schulman68a,Schulman69a,Schulman71a},
\citet{Laidlaw/MoretteDeWitt71a} and \citet{Dowker72a} and
their numerous followers.  Despite its fundamental importance, to my
knowledge, almost all the attempts of justifying the
decomposition~\eqref{eq:Kdecomp} concern the fact that a
unitary representation is \emph{sufficient} to get a consistent model
for the quantum evolution. The only exceptions being the works of
\citet{Laidlaw/MoretteDeWitt71a} and
\citet[\S\,23.3]{Schulman81a}, to which the literature on the
subject seems to always refer eventually. While underlying their major
contributions, I will, in the same time, try to explain why these rely
on unsatisfactory weak points and therefore are, in my opinion,
incomplete and require to be rebuilt. In
section~\ref{sec:crystal} will illustrate some of the previously discussed points in
the more concrete, but still general, models whose non trivial
topology is induced by periodic boundary conditions. Surprisingly,
after the pioneer article on the subject done by
\citet{Schulman69a}, only caricature models seem to have been
retained in the literature whereas the generality and the simplicity
of the decomposition~\eqref{eq:Kdecomp} for spatially periodic models
would deserve a better attention. The concluding section~\ref{sec:homology}
will emphasize the difference that may be put to experimental tests
between the unitary representation of the first homotopy group and
of another topological group like the first homology group.

\section{Context, stakes and starting concepts}\label{sec:intro}

Topology shares a long history with physics since the \textsc{xix}th
century \citep{Nash99a,Mouchet18a}
   and was introduced into the
quantum arena by Dirac's \citeyear{Dirac31a} seminal work on the
magnetic monopole\footnote{Not referring explicitly to topology does
  not mean, of course, that it is absent, all the more so when the
  mathematical concepts were not stabilized: even before Poincar\'e
  works at the turn of the \textsc{xix-xx}th centuries, topological
  arguments irrigated fluid dynamics and electromagnetism thought the
  works of Helmholtz and of the Anglo-Irish-Scottish school including
  Stokes.  In his paper, Dirac follows repeatedly a topological
  reasoning.}.  Within the Schr\"odinger formalism, the topology of
the configuration space mainly appears through the boundary conditions
imposed to the wavefunctions that constitute the Hilbert space of
states. However, in this context,  untangling the global and, by definition, robust
topological properties from the local (differential) ones is not
straightforward; all the more so than, when dealing with a curved
manifold, the definition of the momentum operator and, more generally,
the set up of the quantum canonical formalism is actually far from
being canonical. The long history---initiated also by
\citet{Dirac27a} himself---and the abundant literature, not free
of tough controversies, about the quantum operator associated with an
angle variable (in phase as well as in configuration space) reflects
these concerns\footnote{For an historical survey on the quantum phase
  operator see \citep{Nieto93a} and for a compilation of papers see
  \citep{Barnett/Vaccaro07a}.}. By offering a direct connection between
the quantum evolution and the paths in configuration space, Feynman's
formulation in its original form \citeyearpar{Feynman42a,Feynman48a},
hides better but does not get rid of the ordering-operator ambiguities,
nor does it make disappear the issues that emerge when trying to
associate a quantum transformation to a (non linear) change of
variables required when covering a manifold (different
from~$\RR^\Deg$) with several patches of curvilinear coordinates
(for instance see \citealp[chap.~7]{Pauli73a};
\citealp{DeWitt57a};
\citealp{Edwards/Gulyaev64a};
\citealp{McLaughlin/Schulman71a};
\citealp[for a review]{Dowker74a};
\citealp{Fanelli76a} for phase-space path
integrals;
\citealp{Gervais/Jevicki76a} for
configuration-space path integrals;
\citealp[chap.~24]{Schulman81a}; a more
recent presentation on these matters can be found in
\citealp[chap.~2]{Prokhorov/Shabanov11a}). However, the great advantage of writing the propagator as
the result of a collective interference from a bunch of paths in
configuration space allows to clearly separate the local properties,
that are encapsulated in the Lagrangian, from the topological ones,
that are encapsulated in the global properties of paths on which the
integral is computed. 
 It is worth mentioning that the appealing
formulation of path integrals in terms of phase-space paths, despite
its many assets over the path integrals in configuration space, seems
to be of poor interest when dealing with topological properties: the
reason is mainly that the phase-space paths that mainly contribute to
the propagator, though continuous is position, are discontinuous in
momenta (\citealp[eq.~(50)]{Feynman48a};
\citealp[fig.~7-1]{Feynman/Hibbs65a})
and by definition, discontinuity is ruled out from topological
considerations.

The freedom of choosing the coefficients~$E$ as the images a unitary
representation of the fundamental group of the configuration space
provides extra resources when combined to the freedom of choosing a
Lagrangian alone. It offers a way to include some features and probe
some properties that are insensitive to out-of-control perturbations.
It unifies in a coherent and common scheme the quantum treatment of
gauge models, including the original Dirac monopole but also the
Ehrenberg-Siday-Aharonov-Bohm effect
\citep{Ehrenberg/Siday49a,Aharonov/Bohm59a}.  As far as
non-relativistic particles are concerned, not only it provides a new
understanding on the fundamental dichotomy between bosons in fermions
through the two possible unitary scalar representations of the
permutation group, the trivial one and the signature
\citep{Laidlaw/MoretteDeWitt71a}, but it opens the doors, in effective
two dimensions, to the intermediate behaviour of anyons \cite[for
  instance and other papers in chaps.~5 and 6 of Shapere \& Wilczek's
  collection]{Leinaas/Myrheim77a,Wilczek82a,Arovas89a}
  whose existence has been proven recently by a direct experimental
evidence \citep{Bartolomei+20a}.  These two examples, gauge theory and
the statistics of identical non relativistic particles, not to speak
about solitons in field theory, are sufficient by themselves to
understand the importance of the possibilities brought by the
decomposition~\eqref{eq:Kdecomp}.  However convincing the arguments
that historically led to it, it is worth exploring if there would be
alternatives to the possible choices of~$E$'s and understand
thoroughly, on physical grounds, the reasons why there are not.

All along this work, I have tried to keep the notations to be
self-explaining and standard enough. The reader who has already some
acquaintance with the subject may skip directly to
section~\ref{sec:Kdecomp}, possibly coming back to the following for
clarification.  In the remaining of the present
section~\ref{sec:intro}, all the definitions and notations that will
be used are specified.  The reader is supposed to be familiar with the
elementary notions of homotopy theory that are sketchily provided for
the sake of self-containedness. For more complete constructions,
examples and proofs see the chapters~1 in the remarkable books of
\citet{Hatcher02a} or \citet{Fomenko/Fuchs16a}.
After \citet{Schulman69a,Schulman68a,Schulman71a},
\citet{Dowker72a} was the first to emphasize that a neat
justification of~\eqref{eq:Kdecomp} requires an auxiliary space called
the universal covering space of the configuration space~$Q$, which we
will denote by~$\barQ$. Whereas its definition and its properties have
now became overspread in the physics literature on the subject we deal
with, it seems that its general and systematic \emph{construction}
(therefore the proof of its existence) remains confined to some
algebraic topology textbooks \cite[\S\,1.3 from p.~63 and
  up]{Hatcher02a} or \cite[\S\,6.12]{Fomenko/Fuchs16a}. Therefore, to
understand why~$\barQ$ does not come out of the blue, a special,
somehow extended, place is devoted below to this construction.

\paragraph{Paths and concatenation.} To be more specific, 
 the configuration space of the system with~$\Deg$
degrees of freedom is supposed to be
 a real manifold~$Q$ equipped with a sufficiently smooth differential
structure so that a Schr\"odinger equation (resp. a Lagrangian) can be defined to model the quantum (resp. classical) evolution.
  A path~$\mathscr{C}$ will be a continuous
map~$t\in[t_i,t_f]\mapsto q(t)\in Q$; in addition to its geometrical
image, made of 1d-continuous subset of~$Q$, it is important to keep in
mind that the dynamical course, through the parametrisation in
time~$t$, is an essential characteristic of~$\mathscr{C}$: even though
they share the same image, two paths having a different
velocity~$\dot{q}$ at the same point will be considered as
distinct\footnote{Whereas, in differential geometry, (oriented)
  paths or curves are usually defined up to a (monotonically
  increasing) bijective continuous reparametrisation that can always
  be taken to be, say,~$s\in[0,1]$.  }. To any
path~$\mathscr{C}$ joining~$q_i\DEFt q(t_i)$ to~$q_f\DEFt q(t_f)$ is
associated a unique inverse denoted by~$\mathscr{C}^{-1}$ obtained by
reversing the course of time through the reparametrisation
$\tilde{q}(t)=q(t_f+t_i-t)$. A path~$\mathscr{C}'$ joining~$q'_I\DEFt
q'(t^{}_I)$ to~$q'_f\DEFt q'(t^{}_f)$ can be concatenated to any other
path~$\mathscr{C}$ joining~$q_i\DEFt q(t_i)$ to~$q_I\DEFt
q(t_I)$ provided that~$t_I\in[t_i,t_f]$ and~$q'_I=q^{}_I$; the
result is a path defined by~$t\in[t_i,t_f]$ such
that~$\tilde{q}(t)=q(t)$ when~$t\in[t_i,t^{}_I]$
and~$\tilde{q}(t)=q'(t)$ when~$t\in[t^{}_I,t_f]$ and will be denoted
by~$\mathscr{C}\cc\mathscr{C}'$ (when~$t_f>t_i$, note that the
chronological ordering is chosen to go from left to right). The concatenation is associative:
$(\mathscr{C}\cc\mathscr{C}')\cc\mathscr{C}''=\mathscr{C}\cc(\mathscr{C}'\cc\mathscr{C}'')$
whenever the concatenation of the three paths is possible.  A loop is
a path such that~$q(t_f)=q(t_i)$. As far as I know, all the relevant
configuration spaces in physics are pathwise-connected (any pairs of
points are the endpoints of a path) and so will be~$Q$. 

\begin{figure}[ht]
\begin{center}
\includegraphics[width=\textwidth]{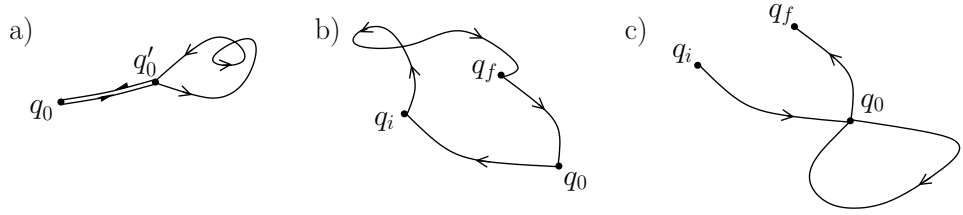}
\caption{\label{fig:homotopie} a)~Every class of paths connecting~$q^{}_0$ to $q'_0$
allows to construct a class of loops  whose basepoint is~$q_0$ from  a class of loops  whose basepoint is~$q'_0$. b)~Every couple of paths connecting~$q_0$ to~$q_i$ and~$q_f$ to~$q_0$ allows to construct a 
class of loops  whose basepoint is~$q_0$
from a class of paths connecting~$q_i$ to~$q_f$ and c)~vice-versa. }
\end{center}
\end{figure}

\paragraph{Homotopy.} Two paths~$q(t)$ and~$q'(t)$ are said to be 
homotopic if they are defined on the same time interval~$[t_i,t_f]$,
if they share the same endpoints, $q'(t_i)=q(t_i)$, $q'(t_f)=q(t_f)$
and if they can be continuously deformed one into the other. This
equivalence relation allows to classify the paths within homotopy
classes that will be denoted by lower-case Gothic letters. When there
is a risk of ambiguity, the endpoints will be specified as
in~$\mathfrak{c}_{q_i,q_f}$.  The set of all classes sharing the same
endpoints will be denoted by~$\pi_1(q_i,q_f)$.  The concatenation law
is transferred to the set of classes: $\mathfrak{c}^{}_{q_i,
  q_I}\cc\mathfrak{c}'_{q_I,q_f}$ is the common homotopic class of
every path~$\mathscr{C}\cc\mathscr{C}'$ obtained for
any~$\mathscr{C}\in\mathfrak{c}^{}_{q_i, q_I}$ and
any~$\mathscr{C}'\in\mathfrak{c}'_{q_I, q_f}$.  This classification
erases the time parametrisation since two different paths~$t\mapsto
q(t)$ and~$s\mapsto q\big(t(s)\big)$ are homotopic whenever~$t(s)$ is
a continuous bijection. We will denote by~$\mathfrak{c}^{-1}_{q_f,
  q_i}$ the homotopy class of~$\mathscr{C}^{-1}$ whatever 
is~$\mathscr{C}\in\mathfrak{c}_{q_i, q_f}$. When restricted to the
set~$\pi_1(q_0,q_0)$ of the classes of loops~$\mathfrak{l}_{q_0,q_0}$,
the concatenation becomes an internal law, having a neutral element,
the class~$\mathfrak{e}_{q_0}$ of all the loops starting and ending at
the basepoint~$q_0$ that are homotopic to the constant path~$q(t)=q_0$
for all~$t$.  Then,~$\mathfrak{l}^{-1}_{q_0, q_0}$ is precisely the
inverse of~$\mathfrak{l}^{}_{q_0, q_0}$ for the concatenation
law. Endowed with the latter, $\pi_1(q_0,q_0)$ is a group for every
choice of the basepoint~$q_0$ and all these groups are isomorphic one
to the other through a left and right concatenation by a class and its
inverse connecting the two basepoints---this is  a particular case of
equation~\eqref{eq:bijpi1} below when~$q_f=q_i$ (Fig.~\ref{fig:homotopie}a)---therefore, they can be
upgraded to an abstract group~$\pi_1(Q)$, independent of~$q_0$, called
the fundamental group of~$Q$ which constitutes a topological invariant
of~$Q$ (\textit{i.e.} preserved by any continuous deformation of~$Q$).
As shown in~Fig.~\ref{fig:2tori}, this group is not necessarily Abelian.
The configuration space~$Q$ is said to be simply-connected when all
its loops can be continuously deformed into one point, in other words
when~$\pi_1(Q)=\{\mathfrak{e}\}$. Otherwise, $Q$~is said to be
multiply-connected.

\begin{figure}[ht]
\begin{center}
\includegraphics[width=\textwidth,clip=0]{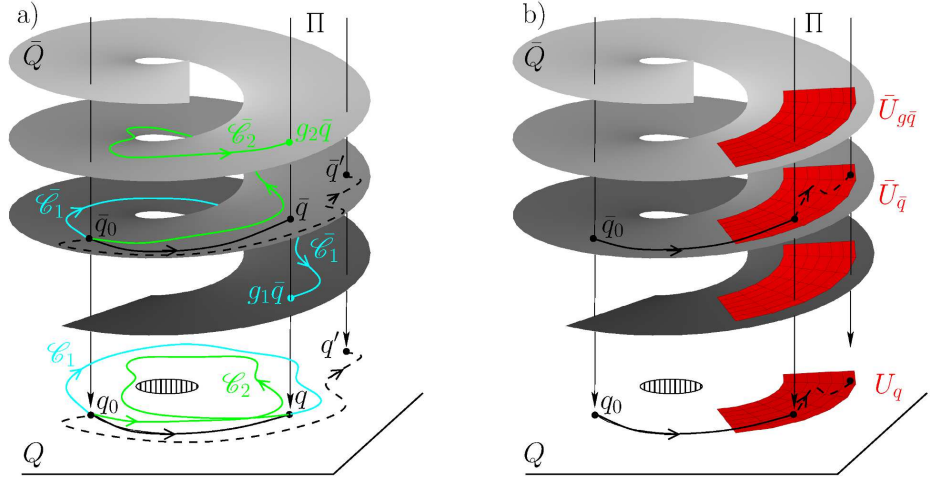}
\caption{\label{fig:QrevetU} (colour on line) a)~Construction of the universal covering space~$\barQ$ from~$Q$ whose
multi-connectedness comes from a forbidden region (hatched region). b)~The differential structure 
of~$\barQ$ is obtained 
through the inverse of the projection~$\Pi$ (represented by vertical downward arrows) which allows to lift the coordinate patches~$U_q$ of
the differential manifold~$Q$.  }
\end{center}
\end{figure}

When~$q_f\neq q_i$, $\pi_1(q_i,q_f)$ is not a group (because
concatenation between two of its elements is not possible) but can be
constructed from~$\pi_1(Q)$ in the following way: for any choice
of~$q_0\in Q$, $\mathfrak{c}_{q_0, q_i}\in\pi_1(q_0,q_i)$
and~$\mathfrak{c}'_{q_0, q_f}\in\pi_1(q_0,q_f)$, each
element~$\mathfrak{c}_{q_i, q_f}\in\pi_1(q_i,q_f)$ has a unique
decomposition of the form (Figs.~\ref{fig:homotopie}b,c)
\begin{equation}\label{eq:bijpi1}
  \mathfrak{c}^{}_{q_i, q_f}=\mathfrak{c}^{-1}_{q_i, q_0}\cc\mathfrak{l}^{}_{q_0,q_0}\cc\mathfrak{c}'_{q_0, q_f}
\end{equation} 
where~$\mathfrak{l}_{q_0,q_0}\in\pi_1(q_0,q_0)$: trivially, $\mathfrak{l}_{q_0,q_0}$ is uniquely given 
by~$\mathfrak{c}^{}_{q_0, q_i}\cc\mathfrak{c}^{}_{q_i, q_f}\cc\mathfrak{c}'^{-1}_{q_f, q_0}$.
We will take advantage of this bijective map between~$\pi_1(q_i,q_f)$ and~$\pi_1(Q)$ to label
the elements of the former with the  elements of the latter.
\paragraph{Construction of the universal covering space.} (Fig.~\ref{fig:QrevetU}a) Once a
basepoint~$q_0$ is chosen in~$Q$ then the universal covering space can be obtained as
\begin{equation}\label{def:Qq0}
  \barQ_{q_0}\DEF\bigcup_{q\in Q}\pi_1(q_0,q)
\end{equation}
which is a disjoint union, that is, for every~$\bar{q}\in\barQ_{q_0}$ 
there exists a unique~$q=\Pi(\bar{q})\in Q$ such 
that~$\bar{q}\in\pi_1(q_0,q)$. Because of the bijective
map between any two~$\pi_1(q_i,q_f)$ obtained from~\eqref{eq:bijpi1},
two different choices of basepoint will provide two bijectively related~$\barQ_{q_0}$'s
and all these sets can be abstracted into a basepoint-independent set~$\barQ$.
 When~$Q$ is simply-connected, all the~$\pi_1(q_0,q)$ have just
one element that can be identified with the endpoint~$q$ itself and
therefore~$\barQ=Q$. When~$Q$ is multiply-connected, $\barQ$~is a
patchwork made of several copies of~$Q$, each being labelled by the
elements of~$\pi_1(Q)$: more precisely, each~$q$ is in correspondence
with several elements in~$\barQ$, namely the elements
of~$\pi_1(q_0,q)$ which are themselves, as we have seen, bijectively
related to~$\pi_1(Q)$. To avoid multivaluedness this correspondence is
rather described by its inverse, the projection~$\Pi$ from~$\barQ$
to~$Q$ defined above, which associates to each
element~$\bar{q}=\mathfrak{c}_{q_0,q}\in\barQ$ the final point~$q$ of
any of the paths in~$\mathfrak{c}_{q_0,q}$. In other
words~$\Pi^{-1}(q)=\pi_1(q_0,q)$.

From  any~$\bar{q}=\mathfrak{c}_{q_0,q}\in\barQ$ and any~$\mathfrak{l}\in\pi_1(q_0,q_0)$ isomorphically associated
to~$g\in\pi_1(Q)$
the class~$\mathfrak{l}\cc\mathfrak{c}_{q_0,q}$ still remains in~$\pi_1(q_0,q)$ and therefore
corresponds to a class~$\bar{q}'=\mathfrak{c}'_{q_0,q}$. Then, to each element~$g\in\pi_1(Q)$
we have a map~$T_g:\bar{q}=\mathfrak{c}_{q_0,q}\mapsto\bar{q}'=\mathfrak{l}\cc\mathfrak{c}_{q_0,q}$ 
that transforms an element of~$\Pi^{-1}(q)$ into another element of~$\Pi^{-1}(q)$.
For convenience we will use the lighter notation~$g\bar{q}\DEFt T_g(\bar{q})$. 
Because of its associativity, the concatenation is isomorphically
transferred to the composition of the~$T$'s: $T_{g'}\circ
T_g=T_{g'g}$.  The~$T$'s define
an action of the fundamental group of~$Q$ on its universal covering
space~$\barQ$. Because~$\mathfrak{l}\cc\mathfrak{c}_{q_0,q}=\mathfrak{c}_{q_0,q}$
if and only if~$\mathfrak{l}=\mathfrak{e}_{q_0}$, the group action is
free that is, by definition, for every~$\bar{q}$ in~$\barQ$,
$g\bar{q}=\bar{q}$ if and only if~$g$ is the neutral
element~$e$ of~$\pi_1(Q)$.

Conversely, from any pair~$\bar{q}'=\mathfrak{c}'_{q_0,q}$
and~$\bar{q}=\mathfrak{c}^{}_{q_0,q}$ there exists a
unique~$g\in\pi_1(Q)$---the one associated to the
loop~$\mathfrak{c}'_{q_0,q}\cc\mathfrak{c}^{-1}_{q,q_0}$
in~$\pi_1(q_0,q_0)$---such that~$g\bar{q}=\bar{q}'$.  One can then
adopt the more common inverse perspective and recover~$Q$
from~$\barQ$: it is the set of the orbits in~$\barQ$ under the action
of~$\pi_1(Q)$ or, in other words, we have~$Q=\barQ/\pi_1(Q)$ the set
of equivalence classes in~$\barQ$ where two elements~$\bar{q}'$ and
$\bar{q}$ are defined to be equivalent if there is a~$g\in\pi_1(Q)$
such that~$\bar{q}'=g\bar{q}$.

\paragraph{The differential structure of the universal covering space.} (Fig.~\ref{fig:QrevetU}b) The 
differential structure of the manifold~$Q$ can be lifted to~$\barQ$
for the main reason that the open sets that cover~$Q$, from which the
charts are defined, can be chosen to be simply-connected. For
any~$\bar{q}=\mathfrak{c}_{q_0,q}\in\barQ$, there exists a
simply-connected open set~$U_q$ in~$Q$ containing~$q$, isomorphic to
an open set in~$\RR^\Deg$.  It can be lifted into~$\barU_{\!\bar{q}}$
defined to be the subset of~$\barQ_{q_0}$ made of all the
classes~$\mathfrak{c}_{q_0,q'}$ such that there exists a path
between~$q$ and~$q'$ entirely included in~$U_q$ or, in other
words~$\bar{q}'=\mathfrak{c}'_{q_0,q'}$ will be in~$\barU_{\!\bar{q}}$
if and only if the
class~$\mathfrak{c}^{-1}_{q^{},q_0}\cc\mathfrak{c}'_{q_0,q'}$ contains a
path included in~$U_q$. This requires of course that~$q'\in
U_q$. Clearly~$\bar{q}\in\barU_{\!\bar{q}}$ because the constant path
equal to~$q$ is in~$U_q$ but~$g\bar{q}\not\in\barU_{\!\bar{q}}$ for
all~$g\neq e$. Indeed, suppose~$\bar{q}'=g\bar{q}$ belongs
to~$\barU_{\!\bar{q}}$
then~$\mathfrak{c}^{-1}_{q^{},q_0}\cc\mathfrak{c}'_{q_0,q'}$ is a class
of loops (because~$q'=q$) and this class would be~$\mathfrak{e}_q$
(because it contains a loop included in~$U_q$ which is
simply-connected). Therefore, we would
have~$\mathfrak{c}'_{q_0,q'}=\mathfrak{c}^{}_{q_0,q^{}}$ that
is~$g\bar{q}=\bar{q}$ and hence, as we have seen above,~$g=e$.  Then,
all the~$\barU_{\!g\bar{q}}$ that can be constructed in the same way
are pairwise disjoint.

Moreover, being a differential manifold, $Q$~is also locally
pathwise-connec\-ted (every neighbourhood of every point contains a
pathwise-connec\-ted neighbourhood). Then, for every~$q'$ belonging
to~$U_q$ there exists a path in~$U_q$ connecting~$q$ and~$q'$. Its
class is uniquely defined because~$U_q$ is simply-connected
and therefore there exists a
unique~$\bar{q}'$ in~$\pi_1(q_0,q')$, given
by~$\mathfrak{c}_{q_0,q}\cc\mathfrak{c}_{q,q'}$, belonging
to~$\barU_{\!\bar{q}}$.
 
Therefore~$\Pi^{-1}(U_q)=\bigcup_{g\in\pi_1(Q)}\barU_{\!g\bar{q}}$
appears to be a disjoint union and each~$\barU_{\!g\bar{q}}$ is
bijectively related to~$U_q$ through the
restriction~$\Pi_{\;\rule[0ex]{.1ex}{1.5ex}\scriptscriptstyle\,\barU_{\!g\bar{q}}}$ which
happens to be a homeomorphism since every neighbourhood of~$q$
included in~$U_q$ can be lifted in an analogous way and can be used to
define a basis of open sets in~$\barQ$.

The composition of these homeomorphisms transfer the charts
covering~$Q$ in charts covering~$\bar{Q}$ which eventually inherits of
all the differential structure of~$Q$.

The class~$\bar{q}_0\DEFt\mathfrak{e}_{q_0}$ is a privileged element
of~$Q_{q_0}$ and  we can safely identify~$\barU_{\!\bar{q}_0}$
with~$U_{q_0}$ by
considering~$\Pi_{\;\rule[0ex]{.1ex}{2ex}\scriptscriptstyle\,\barU_{\!\bar{q}_0}}$ as a
trivial inclusion map.

\paragraph{Lifted paths.} 
Every path~$\mathscr{C}$ in~$Q$ given by~$t\in[t_i,t_f]\mapsto q(t)$ connecting~$q_i$ to~$q_f$, once
a~$\bar{q}_i$ is chosen in~$\Pi^{-1}(q_i)$,
is lifted into a unique path~$\bar{\mathscr{C}}$ in~$\barQ$ given by~$t\in[t_i,t_f]\mapsto\bar{q}(t)$
where~$\bar{q}(t)\in\Pi^{-1}\big(q(t)\big)$ is uniquely defined
by covering~$\mathscr{C}$ with simply-connected patches on which the restriction of~$\Pi^{-1}$ is bijective.
Two non-homotopic paths sharing the same endpoints in~$Q$ will be lifted 
in~$\barQ$ into two paths ending to different~$\bar{q}_f=\bar{q}(t_f)$ if they both start at~$\bar{q}_i$.

\paragraph{Simply-connectedness of the universal covering space.} 
So to speak, $\barQ$~is obtained by unfolding~$Q$ in order to get a
simply connected space. If we consider a loop~$\mathfrak{L}$ in~$\barQ$
given by~$\bar{q}(t)=\mathfrak{c}_{q_0,q(t)}$ such
that~$\bar{q}(t_i)=\bar{q}(t_f)$, then its projection by~$\Pi$ in~$Q$,
namely~$q(t)$, is a loop~$\mathscr{L}$ in~$\mathfrak{e}_{q_i}$
precisely
because~$\mathfrak{c}_{q_0,q(t_i)}=\mathfrak{c}_{q_0,q(t_f)}$.  Then,
there exists a continuous deformation that contracts~$\mathscr{L}$
into the constant path equal to~$q_i$ which can be lifted for
continuously deform the original loop~$\mathfrak{L}$ in~$\barQ$ into the
constant path~$\bar{q}(t)=\bar{q}(t_i)$ for all~$t$, therefore~$\barQ$
is indeed simply connected.

\section{Emergence of the unitary representation of~$\pi_1(Q)$}\label{sec:Kdecomp}

\subsection{General characteristics of the propagator}

All the quantum evolution operators~$\hat{U}(t_f,t_i)$ share the following characteristic properties:
for any times~$(t_i,t,t_f)$, we have the composition law 
\begin{subequations}\label{eq:Uop}
\begin{equation}\label{eq:compositiondeU}
  \hat{U}(t_f,t)\hat{U}(t,t_i)=\hat{U}(t_f,t_i)\;,
\end{equation}
endowed with the neutral element representing a non-evolution
\begin{equation}\label{eq:elementneutreU}
        \hat{U}(t_i,t_i)=1\;,  
\end{equation}
which makes the~$\hat{U}$'s unitary provided the exchange of time arguments corresponds to Hermitian conjugation
\begin{equation}\label{eq:unitariteU}
  \big(\hat{U}(t_f,t_i)\big)^{*}=\hat{U}(t_i,t_f)\;.
\end{equation}
\end{subequations}

The chronological ordering is completely free, in particular one cannot impose systematically~$t\in[t_f,t_i]$ 
because
  conditions~\eqref{eq:Uop} do not allow to conclude that~$\smash{\big(\hat{U}(t_f,t_i)\big)^{*}=\big(\hat{U}(t_f,t_i)\big)^{-1}}$
  if~$\smash{\hat{U}(t,t)}$ cannot be decomposed into~$\smash{\hat{U}(t,t_I)\hat{U}(t_I,t)}$ for any~$t_I$:
 it is required to use the 
  composition law between operators whose arguments change their chronological order one from the other. 
We note also that~\eqref{eq:elementneutreU} 
is not a consequence of~\eqref{eq:compositiondeU} (by taking~$t_f=t_i$ for instance)
if we refrain posing a priori that the~$\hat{U}$'s are invertible. 
The propagators~$K(q_f,t_f,q_i,t_i)$ in configuration space~$Q$ are (generalised) functions of
two points~$(q_f,q_i)$ in~$Q$ that can be thought has the matrix elements~$\bra{q_f}\hat{U}(t_f,t_i)\ket{q_i}$
where~$\{\ket{q}\}$ denotes a non-normalisable basis of the Hilbert space 
on which the~$\hat{U}$'s are defined. The properties~\eqref{eq:Uop} have
their exact translation in terms of propagators \citep[eqs.~30.6,7,8]{Pauli73a}
\begin{subequations}\label{eq:K}
\begin{equation}\label{eq:compositiondeK}
   K(q_f,t_f,q_i,t_i)=\int_Q K(q_f,t_f,q,t) K(q,t,q_i,t_i)\, \dmat^{\Deg} q ;
\end{equation} 
\begin{equation}\label{eq:elementneutreK}
     K(q_f,t_i,q_i,t_i)=\delta(q_f-q_i);
\end{equation}
and
\begin{equation}\label{eq:unitariteK}
  \Big(K(q_f,t_f,q_i,t_i)\Big)^*= K(q_i,t_i,q_f,t_f)\;.
\end{equation}
\end{subequations}
We denote by~$\dmat^{\Deg}q$ a given measure on~$Q$ (including a non-homogeneous Jacobian
when curvilinear coordinates are used) associated with the Dirac function~$\delta$ such 
that~$\int_Q f(q)\delta(q'-q)\, \dmat^{\Deg} q=f(q')$ for any test function~$f$ defined on~$Q$.
It is important to note that the integral that constitutes  the right-hand side of~\eqref{eq:compositiondeK} 
covers the whole~$Q$: this is the reason why,
when dividing the evolution into a sequence of infinitesimal-time slices, to build up, by their
composition, the path integral
\begin{equation}\label{eq:Kpath}
  K(q_f,t_f,q_i,t_i)=\int
\EXP{\frac{\imat}{\hbar}S[\mathscr{C}]}\dmat[\mathscr{C}]\;,
\end{equation}
the integration domain includes all the paths~$\mathscr{C}$ on~$Q$ such that~$q(t_i)=q_i$ and~$q(t_f)=q_f$.  
Since we want to emphasize the topological properties, we ought not to
choose an explicit expression of the action~$S$ and actually we will not. We will retain only
 its additivity by concatenation of paths:
\begin{equation}\label{eq:additiviteS}
  S[\mathscr{C}_2\cc\mathscr{C}_1]=S[\mathscr{C}_2]+S[\mathscr{C}_1];\qquad S[\mathscr{C}^{-1}]=-S[\mathscr{C}]
\end{equation}
(the latter does not assume any time-reversal invariance, which is in
fact not satisfied as soon as a magnetic field is present; rather, it
may provide a definition of the transformed Lagrangian under~$t\mapsto
t_f+t_i-t$).  We will neither use a precise definition of the path
integrals. We shall assume that actually the right-hand side
of~\eqref{eq:Kpath} satisfies~\eqref{eq:K} which not an easy statement
to prove or, conversely, that must be included in any constructivist
approach.  An important departure from the original construction
proposed by \citet[eq.~(4-28)]{Feynman/Hibbs65a} is that their
propagator is defined to be zero for~$t_f<t_i$ or, equivalently, they
consider the matrix elements of~$\hat{U}(t_f,t_i)$ multiplied by the
Heaviside step function~$\Theta(t_f-t_i)$. We will rather not to
because we want to preserve the property~\eqref{eq:unitariteK} which
is essential to the group property of the~$\hat{U}$'s; we will keep working
with a function~$K$ that fulfills the same evolution equation as a
normalisable state, namely the time dependent Schr\"odinger equation,
without any supplementary~$\delta(t-t_i)$ terms turning it into a
retarded Green function. For the same reason, we will not allow to use
the Wick-substitution ``mantra'' that leads to an irreversible evolution governed
by a semi-group. The difficulty of defining mathematically an
oscillatory path integral is the sign that using an imaginary time is
not harmless from the physical point of view; as soon as we suppress,
by construction, the central notion of quantum interferences, one is
expected to miss a lot of physics including the topological phases as
they appear in~\eqref{eq:Kdecomp}. We will show how the latter are
directly connected to the unitarity of the evolution.

For~$t_f\neq t_i$, one expects the function~$(q_f,q_i)\mapsto
K(q_f,t_f,q_i,t_i)$ to be smooth on the configuration space if the
Schr\"odinger time-dependent equation satisfied by~$K$ involves
potentials that are regular enough.  In particular, for a model
invariant under time translations, from a stationary orthonormal
eigenbasis~$\{\ket{\phi_\nu}\}$ labelled by the quantum numbers~$\nu$
and corresponding to an energy spectrum~$\{\epsilon_\nu\}$, we have the
following spectral decomposition of~$K$ in terms of the corresponding
wavefunctions~$\phi_\nu(q)\DEFt\braket{q}{\phi_\nu}$ defined on~$Q$
\begin{equation}\label{eq:Kpsipsi}
  K(q_f,t_f,q_i,t_i)=\sum_\nu \phi^{}_\nu(q_f)\phi^*_\nu(q_i)\,
  \EXP{-\frac{\imat}{\hbar}(t_f-t_i)\epsilon_\nu}
\end{equation}
and the smoothness of~$K$ is at least the same as the smoothness of
the stationary wavefunctions.

\subsection{Propagators in the universal covering space}

Once the domain of integration in the right-hand side of~\eqref{eq:Kpath} has been split in homotopy 
classes~$\mathfrak{c}_{q_i,q_f}$, the weights~$E(\mathfrak{c}_{q_i,q_f})$ 
 must be chosen in order
for the new~$K$ given by~\eqref{eq:Kdecomp} to still satisfy~\eqref{eq:K}
and, then, to keep its interpretation of being 
the density of probability  amplitude for reaching the configuration~$q_f$ at
time~$t_f$
assuming that the system is in the configuration~$q_i$ at~$t_i$.
On the other hand,
the partial path integrals defined by
\begin{equation}\label{def:kc}
  k_{\mathfrak{c}}(q_f,t_f,q_i,t_i)\DEF\int_{\mathscr{C}\in\mathfrak{c}}
\EXP{\frac{\imat}{\hbar}S[\mathscr{C}]}\dmat[\mathscr{C}]
\end{equation} 
are not expected to satisfy~\eqref{eq:K} and, notably, one should
expect that
\begin{equation}\label{eq:kneeqdelta}
  k_{\mathfrak{c}}(q_f,t_i,q_i,t_i)\neq\delta(q_f-q_i)
\end{equation}
because to obtain the right-hand side requires a path integration with
no restriction on the homotopy classes as in~\eqref{eq:Kpath}.  In
fact, one expects the function~$(q_f,q_i)\mapsto
k_{\mathfrak{c}}(q_f,t_f,q_i,t_i)$ to have discontinuities in~$Q$
because there is no way of deforming continuously, say, a class of
paths~$\mathfrak{c}_{q_i,q(s)}$ along a non-trivial loop~$q(s)$
without ending with another class of paths.  For instance, let us
choose~$s\in[0,1]\mapsto q(s)$ to be a loop~$\mathscr{L}$ in~$Q$
with~$q(0)=q(1)=q_f$ which is not continuously deformable into the
constant path~$q_f$. Then, we can try to maintain the continuity
of~$s\mapsto k_{\mathfrak{c}}(q(s),t_f,q_i,t_i)$ by transforming
continuously the domain of integration of~\eqref{def:kc} in the
following way: to define~$\mathfrak{c}_{q_i,q(s+\dmat s)}$, all the
paths in~$\mathfrak{c}_{q_i,q(s)}$ are concatenated with the portion
of~$\mathscr{L}$ between~$q(s)$ and~$q(s+\dmat s)$ while changing the
time parametrisation to keep the paths always defined
for~$[t_i,t_f]$. Nevertheless, at the end of this process, when the
final point~$s=1$ is reached, all the paths have been concatenated
with~$\mathscr{L}$ modulo a time reparametrisation and therefore
belong to the
class~$\mathfrak{c}'_{q_i,q_f}=\mathfrak{c}^{}_{q_i,q_f}\cc\mathfrak{l}^{}_{q_f,q_f}$;
the latter is different from the class we started with at~$s=0$ if and
only if the class~$\mathfrak{l}_{q_f,q_f}$ of~$\mathscr{L}$ is
different from the neutral class~$\mathfrak{e}_{q_f}$. Having two
different integration domains, $k_{\mathfrak{c}}$
and~$k_{\mathfrak{c}'}$ are a priori different and therefore,
even though the limits~$s\to0^+$ and~$s\to1^-$ lead to the same
value~$q(0)=q(1)=q_f$, they give two different values
for~$k_{\mathfrak{c}_{q_i,q(s)}}(q(s),t_f,q_i,t_i)$. Since~$\mathscr{L}$
is arbitrary with at least one point of discontinuity on it, we expect
to have in~$Q$ at least one hypersurface of codimension~1 of points of discontinuity for
each~$k$.

These singularities of the~$k$'s in~$Q$ do not contradict the
regularity of~$(q_f,q_i)\mapsto K(q_f,t_f,q_i,t_i)$ for~$t_f\neq t_i$:
within the sum~\eqref{eq:Kdecomp}, when following the
loop~$\mathscr{L}$ from~$s=0^+$ to~$s=1^{-}$ a permutation occurs
where all terms are swapped one with an other maintaining
the continuity of the whole sum.  One can then even redefine a
continuous family of functions on~$Q$ with the help of a
Heaviside~$\Theta$ functions splitting the two sides of the
hypersurfaces of discontinuities of
the partial path integrals\footnote{If~$k_n(s_0^+)=k_{\sigma(n)}(s_0^-)\neq k_n(s_0^-)$
  for a permutation~$\sigma$ of the discrete labels~$n$, then the
  function~$\tilde{k}_n(s)\DEFt
  k_n(s)\Theta(s-s_0)+k_{\sigma(n)}(s)\Theta(s_0-s)$ is continuous
  at~$s_0$ and one can express the discontinuous functions~$k_n$ with
  the continuous functions~$\tilde{k}_n$: $k_n(s)=
  \tilde{k}_n(s)\Theta(s-s_0)+\tilde{k}_{\sigma^{-1}(n)}(s)\Theta(s_0-s)$. }
but, as we have seen, these functions cannot be associated with the
same homotopy class on the two sides of these surfaces.

To keep following the same homotopy class without dealing with cuts,
the price to be paid is to somehow establish a distinction
between~$q(0)$ and~$q(1)$ by opening the loop~$\mathscr{L}$.  This is
precisely the reason of using the universal covering space~$\barQ$
where the points~$\bar{q}$ are identified with the homotopy
classes~$\mathfrak{c}_{q_0,q}$ with~$q_0$ being fixed and~$q$ running
through~$Q$ (the definition of~$\barQ$ when attached to~$q_0$ is
given by~\eqref{def:Qq0} and its properties are recalled in
section~\ref{sec:intro}): the homotopic distinction between paths
in~$Q$ (sharing the same endpoints) is transferred to a difference in
the endpoints in the lifted paths in~$\barQ$\footnote{The same line of thought leads to the construction
of the Riemann surfaces from the complex plane. The latter is unfolded into~$n$ connected sheets
to avoid the line cuts of the function~$z\mapsto z^{1/n}$.}.

Let us first convert the sum over~$\pi_1(q_i,q_f)$, the class of paths
in~$Q$ connecting~$q_i$ to~$q_f$, to a sum over the fundamental
group~$\pi_1(Q)$ identified with~$\pi_1(q_f,q_f)$.  By selecting
one~$\mathfrak{c}_0\in\pi_1(q_i,q_f)$,~\eqref{eq:Kdecomp} reads
\begin{equation}
   K(q_f,t_f,q_i,t_i)=\sum_{\mathfrak{l}\in\pi_1(q_f,q_f)}E(\mathfrak{c}_0\cc\mathfrak{l})\int_{\mathscr{C}\in\mathfrak{c}_0\cc\mathfrak{l}}
\EXP{\frac{\imat}{\hbar}S[\mathscr{C}]}\dmat[\mathscr{C}]\;.
\end{equation} 
Then, pick up one~$\bar{q}_i\in\Pi^{-1}(q_i)$, set~$q_0=q_f$ and
define~$\bar{q}_f\DEFt\mathfrak{e}_{q_f}$. Then each path~$\mathscr{C}$
in~$Q$ connecting~$q_i$ to~$q_f$ is lifted into a unique path~$\bar{\mathscr{C}}$ in~$\barQ$ 
connecting~$\bar{q}_i$ to~$g\bar{q}_f$ for~$g$ in~$\pi_1(Q)$ associated to~$\mathfrak{l}$.
When restricted to simply-connected patches on~$Q$,  all the 
differential structure of~$Q$ can be lifted to~$\barQ$, in particular the coordinates charts,
the
action functional and the measure on paths; the definition of which are part of the
 translation of the partial path integral
into the universal covering space according to
\begin{equation}\label{eq:kK}
\begin{split}
  k_{\mathfrak{c}_0\cc\mathfrak{l}}(q_f,t_f,q_i,t_i)&\DEF\int_{\mathscr{C}\in\mathfrak{c}_0\cc\mathfrak{l}}
\EXP{\frac{\imat}{\hbar}S[\mathscr{C}]}\dmat[\mathscr{C}];\\&=
\int_{\bar{\mathscr{C}}\in\bar{\mathfrak{c}}_{\bar{q}_i,g\bar{q}_f}}
\EXP{\frac{\imat}{\hbar}\bar{S}[\bar{\mathscr{C}}]}\bar{\dmat}[\bar{\mathscr{C}}]\DEF\barK(g\bar{q}_f,t_f,\bar{q}_i,t_i)
\end{split}
\end{equation} 
where the integration domain is now the homotopy class~$\bar{\mathfrak{c}}_{\bar{q}_i,g\bar{q}_f}$
of the paths~$\bar{q}(t)$ in~$\barQ$ such that~$\bar{q}(t_i)=\bar{q}_i$ and~$\bar{q}(t_f)=g\bar{q}_f$.
In working in the universal covering space, we have expressed each partial path integrals~$k$
 into a plain
Feynman integral~$\barK(g\bar{q}_f,t_f,\bar{q}_i,t_i)$
 where all the paths connecting two points are considered with no restriction
on their homotopy classes since, by construction, $\barQ$ is simply connected.
The equality~\eqref{eq:kK} concerns the \emph{values}
 of~$k$ and~$\barK$
and not the functions themselves whose arguments are defined in different spaces and whose
smoothness properties are not the same.

\subsection{Linear independence of the~$k$'s}

As any plain Feynman integral, $\barK$ satisfies~\eqref{eq:elementneutreK}, 
\begin{equation}\label{eq:enKbar}
  \barK(g\bar{q}_f,t_i,\bar{q}_i,t_i)=\delta(g\bar{q}_f-\bar{q}_i)\;,
\end{equation}
in contrast with~\eqref{eq:kneeqdelta}. 
Then, if for one reason or another, a linear combination of the form~$\sum_{g\in\pi_1(Q)}A(g)\barK(g\bar{q}_f,t_f,\bar{q}_i,t_i)$ vanishes identically for all times,  by taking~$t_f=t_i$, this implies that the coefficients~$A(g)$ are zero because~$g\bar{q}_f\neq\bar{q}_f$ as soon as~$g\neq e$. From~\eqref{eq:kK}, 
this linear independence of the~$\delta(g\bar{q}_f-\bar{q}_i)$ for~$g\in\pi_1(Q)$
is directly transmitted to the~$k$'s:
\begin{equation}\label{eq:linindk}
  \sum_{\mathfrak{l}\in\pi_1(q_f,q_f)}A(\mathfrak{c}_0\cc\mathfrak{l})k_{\mathfrak{c}_0\cc\mathfrak{l}}(q_f,t_f,q_i,t_i)=0\quad\Longleftrightarrow\quad \forall\mathfrak{l}\in\pi_1(q_f,q_f),\ \ A(\mathfrak{c}_0\cc\mathfrak{l})=0. 
\end{equation}
In other words, we have established that the decomposition~\eqref{eq:Kdecomp} of a given propagator~$K$
is necessarily unique.   

\subsection{Composition}

For the decomposition~\eqref{eq:Kdecomp} to be consistent with~\eqref{eq:compositiondeK}, we must have 
\begin{multline}\label{eq:prodintcheminc1c2}
  \sum_{\mathfrak{c}\in\pi_1(q_i,q_f)} E(\mathfrak{c}) 
\int_{\mathscr{C}\in\mathfrak{c}}\!\!\EXP{\frac{\imat}{\hbar}S[\mathscr{C}]}\dmat[\mathscr{C}]\\=\int_Q\, \dmat^\Deg q_I
\sum_{\substack{\mathfrak{c}_1\in\pi_1(q_i,q_I)\\ \mathfrak{c}_2\in\pi_1(q_I,q_f)}} 
E(\mathfrak{c}_2)\,E(\mathfrak{c}_1)
 \int_{\substack{\mathscr{C}_1\in\mathfrak{c}_1\\\mathscr{C}_2\in\mathfrak{c}_2}}\!\!
\EXP{\frac{\imat}{\hbar}(S[\mathscr{C}_2]+S[\mathscr{C}_1])}\dmat[\mathscr{C}_2]\,\dmat[\mathscr{C}_1]\;.
\end{multline}

Take~$t_i<t_f$ choose~$t_I\in]t_i,t_f[$. 
Every path~$\mathscr{C}$ involved in the integral of the left-hand side, connecting~$(q_i, t_i)$ to~$(q_f, t_f)$, is uniquely obtained by concatenation of one path~$\mathscr{C}_2$ connecting~$(q_I, t_I)$ to~$(q_f, t_f)$
to one path~$\mathscr{C}_1$ connecting~$(q_i, t_i)$ to~$(q_I, t_I)$
where~$q_I$ given by~$q(t_I)$;  the class~$\mathfrak{c}$ of $\mathscr{C}$
is then uniquely decomposed into~$\mathfrak{c}_1\cc\mathfrak{c}_2$ where~$\mathfrak{c}_1$ (resp.~$\mathfrak{c}_2$)
denotes the homotopy class of~$\mathscr{C}_1$ (resp.~$\mathscr{C}_2$). Therefore each path
of the left-hand side appears once and only once among the paths in the  right-hand side.

Conversely, every path involved in the  right-hand side
is obtained by concatenation of a path connecting~$(q_I, t_I)$ to~$(q_f, t_f)$
to a path connecting~$(q_i, t_i)$ to~$(q_I, t_I)$  for a given~$q_I$ and then appears once and
only once in the left-hand side. 

Moreover, because of the additivity property~\eqref{eq:additiviteS}, we can collect the paths of the right-hand side according to
\begin{multline}
  \int_Q\, \dmat^\Deg q_I\sum_{\substack{\mathfrak{c}_1\in\pi_1(q_i,q_I)\\\mathfrak{c}_2\in\pi_1(q_I,q_f)}} E(\mathfrak{c}_2)E(\mathfrak{c}_1)
 \int_{\substack{\mathscr{C}_1\in\mathfrak{c}_1\\\mathscr{C}_2\in\mathfrak{c}_2}}
\EXP{\frac{\imat}{\hbar}(S[\mathscr{C}_2]+S[\mathscr{C}_1])}\dmat[\mathscr{C}_2]\dmat[\mathscr{C}_1]\\
= \sum_{\substack{\mathfrak{c}\in\pi_1(q_i,q_f)\\
\mathfrak{c}=\mathfrak{c}_1\cc\mathfrak{c}_2}} E(\mathfrak{c}_2)E(\mathfrak{c}_1)
 \int_{\mathscr{C}\in\mathfrak{c}}\EXP{\frac{\imat}{\hbar}S[\mathscr{C}]}\dmat[\mathscr{C}]\;.
\end{multline}
The identification with the left-hand side~\eqref{eq:prodintcheminc1c2} reads
\begin{equation}\sum_{\substack{\mathfrak{c}\\
\mathfrak{c}=\mathfrak{c}_1\cc\mathfrak{c}_2}} E(\mathfrak{c})\,
  k_{\mathfrak{c}}(q_f,t_f,q_i,t_i)=\sum_{\substack{\mathfrak{c}\\
\mathfrak{c}=\mathfrak{c}_1\cc\mathfrak{c}_2}} E(\mathfrak{c}_2)\,E(\mathfrak{c}_1)\,k_{\mathfrak{c}}(q_f,t_f,q_i,t_i)
\end{equation}
that is
\begin{equation}\label{eq:Emorphismelacets}
  E(\mathfrak{c}_1\cc\mathfrak{c}_2)= E(\mathfrak{c}_2)\,E(\mathfrak{c}_1)
\end{equation}
because of the linear independence of the~$k_\mathfrak{c}$'s.  Then,
the generalisation~\eqref{eq:Kdecomp} preserves the original
\nompropre{Feynman}'s interpretation: the probability amplitude
brought by the path~$\mathscr{C}=\mathscr{C}_1\cc\mathscr{C}_2$ to the
propagator~$K$ remains equal to the product of the amplitudes brought
by~$\mathscr{C}_1$ and~$\mathscr{C}_2$; since the integral involves
all the possible paths, the two concatenated pieces are considered to
be independent as soon as the continuity of~$\mathscr{C}$ is
maintained.  This multiplication of the amplitudes
reads~$E(\mathfrak{c}_1\cc\mathfrak{c}_2)\,
\EXP{\frac{\imat}{\hbar}S[\mathscr{C}]}=E(\mathfrak{c}_2)\,E(\mathfrak{c}_1)\,
\EXP{\frac{\imat}{\hbar}(S[\mathscr{C}_2]+S[\mathscr{C}_1])}$ which is
guaranteed both by~\eqref{eq:Emorphismelacets} and by the additivity
of the action with respect to concatenation.

As an immediate consequence of~\eqref{eq:Emorphismelacets} by taking for~$\mathfrak{c}_2$ any neutral
class~$\mathfrak{e}_q$ and~$\mathfrak{c}_1\in\pi_1(q_i,q)$ we get
\begin{equation}\label{eq:Eeun}
  E(\mathfrak{e}_q)=1
\end{equation}
and by choosing~$\mathfrak{c}^{}_2=\mathfrak{c}_1^{-1}$,
\begin{equation}\label{eq:Ecm1}
   E(\mathfrak{c}^{-1})= \big(E(\mathfrak{c})\big)^{-1}\;.
\end{equation}

\subsection{Conjugation}

The third and last characteristic property of a propagator is the Hermitian conjugation rule~\eqref{eq:unitariteK}. Then we must have
\begin{equation}
   \sum_{\mathfrak{c}\in\pi_1(q_i,q_f)} E(\mathfrak{c})
\int_{\mathscr{C}\in\mathfrak{c}}\EXP{\frac{\imat}{\hbar}S[\mathscr{C}]}\dmat[\mathscr{C}]
= \sum_{\mathfrak{c}^{-1}\in\pi_1(q_f,q_i)} \big(E(\mathfrak{c}^{-1})\big)^*
\int_{\mathscr{C}\in\mathfrak{c}^{-1}}\EXP{-\frac{\imat}{\hbar}S[\mathscr{C}]}\dmat[\mathscr{C}]
\end{equation}
where all the classes~$\mathfrak{c}^{-1}$ involved in the right-hand side are made of paths connecting~$(q_f, t_f)$ to~$(q_i, t_i)$. Yet, to each of these paths is associated a unique inverse~$\mathscr{C}^{-1}$ connecting~$(q_i, t_i)$ to~$(q_f, t_f)$ whose action is opposite by virtue of~\eqref{eq:additiviteS}:~$S[\mathscr{C}^{-1}]=-S[\mathscr{C}]$. On the right-hand side, the sum on the path in~$\mathfrak{c}^{-1}$ can be obtained
by a sum on the opposite paths in the classes~$\mathfrak{c}$: 
\begin{equation}
  \int_{\mathscr{C}\in\mathfrak{c}^{-1}}\EXP{-\frac{\imat}{\hbar}S[\mathscr{C}]}\dmat[\mathscr{C}]
=\int_{\mathscr{C}\in\mathfrak{c}}\EXP{\frac{\imat}{\hbar}S[\mathscr{C}]}\dmat[\mathscr{C}]
\end{equation}
(if the path integral is defined as a limit of a discretisation, the Jacobian of such a transformation
equals to one because it just consists in a permutation of the discrete coordinates; in 
 a constructivist perspective this Jacobian is defined to be one).
Therefore we obtain
\begin{equation}
   \sum_{\mathfrak{c}} E(\mathfrak{c})\,k_{\mathfrak{c}}(q_f,t_f,q_i,t_i)=
 \sum_{\mathfrak{c}} \big(E(\mathfrak{c}^{-1})\big)^*k_{\mathfrak{c}}(q_f,t_f,q_i,t_i)
\end{equation}
that is
\begin{equation}\label{eq:conjugaisonEcm1}
  E(\mathfrak{c}^{-1})=\big(E(\mathfrak{c})\big)^*\;,
\end{equation}
by using the linear independence of the~$k_{\mathfrak{c}}$'s again.
Combined with~\eqref{eq:Ecm1} we obtain
\begin{equation}\label{eq:Ecs}
  \big(E(\mathfrak{c})\big)^*=\big(E(\mathfrak{c})\big)^{-1}\;.
\end{equation}

\subsection{Unitary representation} 

To turn~$E$ into a  morphism of groups, besides~\eqref{eq:Emorphismelacets}, one must restrict
its arguments to the class of loops~$\pi_1(Q)$. But this can be done by picking up one~$q_0\in Q$
and two
classes~$\mathfrak{c}_f\in\pi_{q_0,q_f}$, $\mathfrak{c}_i\in\pi_{q_0,q_i}$ and use the loops in~$\pi_1(q_0,q_0)$ to label the paths~$\mathfrak{c}$ in the sum~\eqref{eq:Kdecomp}:
\begin{subequations}\label{eq:Kdecompbis}
\begin{align}
  K(q_f,t_f,q_i,t_i)&=\hspace{-1ex}\sum_{\mathfrak{l}\in\pi_1(q_0,q_0)}\hspace{-1ex}E(\mathfrak{c}^{-1}_i\cc\mathfrak{l}\cc\mathfrak{c}^{}_f)\int_{\mathscr{C}\in\mathfrak{c}^{-1}_i\cc\mathfrak{l}\cc\mathfrak{c}^{}_f}
\EXP{\frac{\imat}{\hbar}S[\mathscr{C}]}\dmat[\mathscr{C}];\\
&=
E(\mathfrak{c}^{}_f)\hspace{-1ex}\sum_{\mathfrak{l}\in\pi_1(q_0,q_0)}\hspace{-1ex}E(\mathfrak{l})\int_{\mathscr{C}\in\mathfrak{c}^{-1}_i\cc\mathfrak{l}\cc\mathfrak{c}^{}_f}
\EXP{\frac{\imat}{\hbar}S[\mathscr{C}]}\dmat[\mathscr{C}]\big(E(\mathfrak{c}_i)\big)^{-1};
\end{align}  
\end{subequations}
where now, together with~\eqref{eq:Emorphismelacets} and~\eqref{eq:Ecs},
 the coefficients~$E(\mathfrak{l})$ are the images of a unitary representation of~$\pi_1(Q)$. The pre-
and post-factors~$E(\mathfrak{c}^{}_f)$ and~$\big(E(\mathfrak{c}_i)\big)^{-1}$ warrant that the composition law
is satisfied. 

By the way, in all of the above, we had no need to work with scalar propagators 
exclusively \citep[\S\,5]{Horvathy+89a}.
Where some discrete  quantum numbers~$\alpha$ label
 the components of the wavefunctions to take into account
the spin or some internal degree of freedom of a bounded system, the propagators~$K$ are implicitly
labelled by two such labels, having a matrix-like structure explicitly given by~$K_{\alpha'\alpha}$.
The action functional a priori depends on these numbers and so the endpoints and the position 
kets~$\ket{q,\alpha}$ come with such multiplicity. 
The right-hand side of~\eqref{eq:elementneutreK} implicitly contains a 
Kronecker symbol~$\delta_{\alpha'\alpha}$ and so on. 
The coefficients~$E$ are then given by a matrix whose entries are explicitly~$(E_{\alpha'\alpha})$.
The relation~\eqref{eq:Emorphismelacets} is to be understood as a matrix product and in~\eqref{eq:Ecs}
a Hermitian conjugation is involved. This offers a direct bridge leading to non-Abelian gauge theories \citep{Oh+88a,Balachandran89a}. 

When dealing with scalar wavefunctions and propagators, from~\eqref{eq:Ecs} we deduce
 that each~$E(\mathfrak{c})$ is a pure phase factor and even though~$\pi_1(Q)$ is not commutative,
its~$\mathrm{U}(1)$-representations are. As noted by \citep[eq.~(2.5)]{Horvathy+89a}, 
the group which is then represented,
obtained by quotienting by the non-commutative part of~$\pi_1(Q)$, appears to be the first 
holonomy group (a coarser topological invariant of~$Q$). In that case, the difference of
phases associated with~$E(\mathfrak{c}^{}_f)\big(E(\mathfrak{c}_i)\big)^{-1}$ can be absorbed
by adding a total derivative in the action~$S$. 

\section{Critical discussion on previous arguments}\label{sec:SLdWB}

\begin{figure}[ht]
\begin{center}
\includegraphics[width=\textwidth]{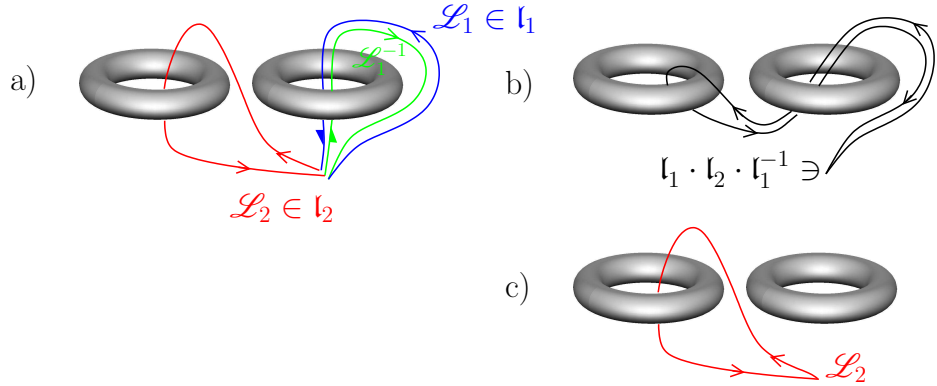}
\caption{\label{fig:2tori}When the interior of two distinct tori are
  removed, we obtain a 3d-space~$Q$ whose fundamental group is not
  commutative.  The
  concatenation~$\mathscr{L}_1\cc\mathscr{L}_2\cc\mathscr{L}^{-1}_1$
of the three loops
  shown in a) leads to a path shown in~b) that cannot be continuously
  deformed into~$\mathscr{L}_2$ in c).  The universal covering
  space~$\barQ$ of~$Q$ is therefore not 2d-crystal-like with a
  periodic structure because the translation group of the latter is
  commutative [to get a visual intuition of~$\barQ$ with the
  same~$\pi_1(Q)$ but in one dimension, see the infinite tree (no loop can appear in a simply-connected graph) in
  \protect\citep[figure p.~59 in~\S\,2.3]{Hatcher02a} 
or \protect\citep[fig.~30 p.~74 of example~5 in \S\,6.9]{Fomenko/Fuchs16a}]. However, such a
 situation could be relevant experimentally by using two (or more)
  shielded ferromagnetic tori like the one used by
  \protect\citet{Tonomura05a} in his experiments on the
  Ehrenberg-Siday-Aharonov-Bohm effect or, in mesoscopic physics, by connecting two conducting tori
  like the gold ring used in~\protect\citep{Webb+85a} provided coherence is maintained all along.  }
\end{center}
\end{figure}

The decomposition~\eqref{eq:Kdecomp} was first proposed by \citet{Schulman68a} for~$Q$ being the
configuration space of a rotating solid (in 2 and 3 dimensions). Whereas this article
focuses on the path integral approach, as its title highlights it, two subsequent
articles \citep{Schulman69a,Schulman71a} explore further some
decomposition of the propagator to other systems but without recourse to path integrals.
The role of the universal covering space is put forward specially in (\citealp[\S\,3 and fig.~1]{Schulman68a}; \citeyear[sec.~I, fig.~3]{Schulman71a}) and occupies a central place in the unified treatment
proposed by~\citet{Dowker72a}. However, all these works, which involve only scalar wavefunctions,
 suppose \emph{a priori} that
the coefficients~$E$ are of unit modulus; this is also taken for granted---and even
so~\eqref{eq:Emorphismelacets} occasionally---in succeeding
articles until recently 
(\citealp[eq.~(1.1)]{Berg81a}; 
\citealp[eq.~(1.1)]{Tarski82a};  
\citealp[eq.~(1)]{Anderson88a}; 
\citealp[eq.~(2.1)]{Horvathy+89a};
\citealp[eq.~(3)]{Ho/Morgan96a}; 
\citealp[before (2.8)]{Tanimura/Tsutsui97a}; 
\citealp[just after eq.~(22)]{Forte05a};
\citealp{Kocabova/Stovicek08a}%
).
 The origin of this hypothesis is easily understood
if one thinks that the covering space has a genuine physical meaning, that is, on which 
wavefunctions have the usual quantum interpretation. Actually, in all the models presented
in these series of papers, and in section~\ref{sec:crystal} as well, $\barQ$~is the primary physical configuration space from which
the multiple-connected base space~$Q$ is built by imposing some boundary conditions (periodicity,
forbidden region). Together with this folding of~$\barQ$ taken to be~$\RR^\Deg$,
the 
wavefunctions are folded as well by identifying~$\bar{\phi}(g\bar{q})$ with~$\bar{\phi}(\bar{q})$
up to a phase~$E(g)=\EXP{\imat\chi(g)}$
 because the latter is unobservable in~$\barQ$:
\begin{equation}\label{eq:psigq}
  \bar{\phi}(g\bar{q})=\EXP{\imat\chi(g)}\bar{\phi}(\bar{q})\;.
\end{equation}
Then, since~$Q$ appears somehow secondary or at last artificially introduced, one has no qualms
about violating the very principles of quantum theory by considering multivalued wavefunctions or propagators in~$Q$\footnote{In the context of the Ehrenberg-Siday-Aharonov-Bohm effect,
see~\citet{Berry80a}'s fair denunciation of the use of multivalued wavefunctions.  }; $\barQ$ never ceases to be
 the genuine physical space where
wavefunctions and propagators remain monovalued. The point of view  adopted
in the present paper is quite the opposite and the ambiguity inherent to some multivalued 
quantities has never been introduced neither in~$Q$ nor, of course, in~$\barQ$. By laying our foundations
on the Feynman path integral, we keep the possibility of considering the multi-connected space~$Q$
as our primary physical space whereas~$\barQ$ is therefore constructed as an auxiliary space to establish the
linear independence~\eqref{eq:linindk} of the partial path integrals. This is not
an undue theoretical issue to consider models where~$\barQ$ cannot pretend to have a physical meaning.
In a Young interference configuration with charged particles, for instance,
 two magnetic 
impenetrable tori---like the one used
in \citet{Tonomura05a}'s famous experiment on the Ehrenberg-Siday-Aharonov-Bohm effect---the
 non-commutativity of the first homotopy group (Fig.~\ref{fig:2tori})
 gives to~$\barQ$ the structure of an infinite
tree-like manifold; such an ``unnatural'' covering space can also be obtained in lower dimension
by considering 8-shaped wire. Clearly, in such situations, the physical preseance must be given to~$Q$
over~$\barQ$ and multivalued quantities in~$Q$ cannot be supported.
In any case, banishing multivalued functions preserves the flexibility of
interpreting $Q$ or~$\barQ$ as the primary physical space.

The first attempt to prove that, in the scalar case, the~$E$'s not
only can but must be obtained from a $\mathrm{U}(1)$-representation of
the first homotopy group~$\pi_1(Q)$ was proposed by
\citet{Laidlaw/MoretteDeWitt71a} in the first part of their
article. There, the linear independence of the partial path integrals
together with their behaviour at~$t_f\to t_i$ was already understood
to be key in determining the weight factors~$E$. Unfortunately, their
arguments suffer from several flaws coming from the ubiquitous
confusion between~$Q$ and~$\barQ$\footnote{Supposedly, this motivated \citet[Introduction]{Dowker72a}
``to present
a somewhat neater and more attractive derivation of [the result~\eqref{eq:Kdecomp}].''}. To prove~\eqref{eq:linindk}, it
seems to be appealing to avoid passing by the universal covering space
but, as far as I know, this challenge remains to be met if ever it
makes sense; in the introductory section~\ref{sec:intro}, I have recalled through~\eqref{def:Qq0} the construction
of~$\barQ$ to show how inseparable it is from the analysis of the
topology of paths in~$Q$. In a subsequent review of which
Morette-DeWitt is also a co-author \citep[p.~295]{DeWittMorette+79a},
it is still written that ``There are two equivalent ways of giving
meaning to~[eq.~(1)]. We give here the one which does not require
auxiliary concepts; the other one \citep{Dowker72a} proceeds via the
universal covering.'' and the proof of~\eqref{eq:linindk} is referred
to~\citep{Laidlaw/MoretteDeWitt71a}. In a more recent mathematical synthesis,
\citet[\S\,II-4, p.~2268]{Cartier/DeWittMorette95a} eventually adopt
\citet{Dowker72a}'s approach and work starting with the universal
covering space together with the hypothesis~\eqref{eq:psigq}.

Coming back to the arguments used  in~\citep{Laidlaw/MoretteDeWitt71a}, their key step~II concerning
the short-time behaviour of~$k_{\mathfrak{c}}$  relies on the debatable assumption that the action is an
increasing function of the length of the (not necessarily classical) path for short-time 
intervals\footnote{To reuse their notations, they write p.~1376 that 
if a path is given by the concatenation~$q(a,a')=q(a,b)q(b,a')$
hence~$S[q(a,a')]>S[q(a,b)]$ or, translated into the notations of the present article, $\mathscr{C}=\mathscr{C}_1\cc\mathscr{C}_2\Rightarrow S[\mathscr{C}]>S[\mathscr{C}_1]$.}.
 This may be true for a quasi-free (in the absence
of vector/scalar potential), stationary, 
short-length path but not in general for the non-infinitesimally short paths they also consider. 
The neighbourhood of hard wall boundaries, where some
diffractive non classical paths may minimise or maximise the action
seem to fall out the scope
of their analysis. In fact, if we want to go on with their semiclassical arguments,
we must take care of the non-commutative limits~$\hbar\to0$ and~$t_f-t_i\to0^+$ and, when~$q_f\neq q_i$ the 
behaviour of~$k_{\mathfrak{c}}(q_f,t_f,q_i,t_i)$ is given by one or more oscillatory
integral for each homotopy class whose prefactor must be tamed  and within this
``battle of exponentials'' between the semiclassical contributions, it is not simple
to identify a winner if there is any. In any case, deducing that~$|E|=1$ exactly
rather that approximately without, say, any real exponential prefactor is, in my opinion, the privilege
of a too restrictive class of models.  
More generally, as argued above, the details of the
differential structure  of the action, should not be relevant in a topological analysis. 

Another objection may be raised when
considering~\citealp{Laidlaw/MoretteDeWitt71a}'s definition of linear
independence.  They use a much stronger condition
than~$\sum_{\mathfrak{l}\in\pi_1}A(\mathfrak{c}_0\cc\mathfrak{l})k_{\mathfrak{c}_0\cc\mathfrak{l}}\equiv0$;
they require that this cancellation should occur for any alternative
choice of~$\mathfrak{c}_0$ while not affecting the coefficients~$A$,
in other words, to transcript their condition (p.~1376, top right
column):
$\sum_{\mathfrak{l}\in\pi_1}A(\mathfrak{c}'_0\cc\mathfrak{l})k_{\mathfrak{c}'_0\cc\mathfrak{l}}\equiv0$
for all~$\mathfrak{c}'_0$
while~$A(\mathfrak{c}'_0\cc\mathfrak{l})=A(\mathfrak{c}_0\cc\mathfrak{l})$. But the latter condition
is not generally fulfilled precisely because a phase factor is allowed to appear
when passing from~$A(\mathfrak{c}'_0\cc\mathfrak{l})$
to~$A(\mathfrak{c}_0\cc\mathfrak{l})$ when~$\mathfrak{c}'_0\cdot\mathfrak{c}^{-1}_0\neq\mathfrak{e}_{q_i}$.

Another criticism, which has no serious repercussion on their argument but is crucial in 
\citet[\S\,23.3]{Schulman81a}'s justification,  
can be brought
when they work with a propagator~$K$ whose value
depends, up to a phase, on a choice of ``mesh'' to label the classes connecting~$q_f$ to~$q_i$ with the 
loops which correspond to our~$\mathfrak{c}_f$ and~$\mathfrak{c}_i$  in~\eqref{eq:Kdecompbis}.
The complete propagator~$K$ cannot depend on the purely conventional choice of~$\mathfrak{c}_f$ and~$\mathfrak{c}_i$
and, therefore, a change of the latter
cannot have any impact on~$K$, even by simply changing its global phase.
If this were the case, we would be led again to the spurious multivalued propagator and then to the not less
spurious multivalued wavefunctions~$\phi(q,t_f)=\int_Q K(q,t_f,q_i,t_i)\phi(q_i,t_i)\,\dmat^\Deg q_i$.
From our starting expression~\eqref{eq:Kdecomp} where no choice of~$(\mathfrak{c}_f,\mathfrak{c}_i)$
is required, or by a direct
elementary computation of the right-hand side of~\eqref{eq:Kdecompbis}, $K(q,t_f,q_i,t_i)$ 
remains completely insensitive to the choice of~$(\mathfrak{c}_f,\mathfrak{c}_i)$.

From a birds-eye view, all the proofs  are suspicious 
that do not use, in one way or another, the characteristic 
property~\eqref{eq:unitariteU}, or, in other words, that do not fully 
use the unitary character of the quantum evolution; indeed, when the first homotopy group
has an infinite number of elements\footnote{When~$\pi_1(Q)$ is finite, \eqref{eq:Emorphismelacets}
is sufficient for the~$E$'s to be given by
the roots of unity. } labelled by some integers~$n$, an exponential~$E_n=\EXP{n\theta}$ with~$\mathrm{Re}\,\theta\neq0$
 should be considered
(their exponentially increase when~$n\to\pm\infty$ may be dominated by the exponential decrease of the corresponding
 oscillatory path integral and the sum~\eqref{eq:Kdecomp} could remain convergent).

\section{Crystalline systems}\label{sec:crystal}

Let us illustrate some of the points raised above in the case of crystals. We will consider a quantum
system made, for simplicity, of one particle (this restriction is not essential)
  whose dynamics is governed by a time-independent 
Hamiltonian expressed in terms of canonical Hermitian operators~$\hat{H}=H(\hat{\tv{p}},\hat{\tv{r}})$ which is spatially periodic on a Bravais lattice~$\mathcal{R}$. Then~$\Deg=\Dim$ the dimension of the direct
space identified with~$\RR^\Dim$. We will denote
by~$\mathbf{R}$ the vectors with integer components that constitute~$\mathcal{R}$.
The unitary operator~$\hat{T}(\mathbf{R})\DEFt\EXP{\imat \hat{\tv{p}}\cdot \tv{R}/\hbar }$
 represents the spatial translation by~$\mathbf{R}$.
All the~$\hat{T}$'s commute one with the other and with~$\hat{H}$. We can diagonalise them 
in the same orthonormal basis \citep{Zak67a}
\begin{subequations}
\begin{align}
  &\hat{H}\ket{\phi_\sigma(\tv{k})}=E_\sigma(\tv{k})\ket{\phi_\sigma(\tv{k})};\\
&\hat{T}(\mathbf{R})\ket{\phi_\sigma(\tv{k})}=\EXP{\imat\tv{k}\cdot\tv{R}}\ket{\phi_\sigma(\tv{k})}
\label{eq:TRpsi}
\end{align}
\end{subequations}
where~$\sigma$ denotes a  set of discrete quantum numbers and~$\tv{k}$  denotes~$\Dim$ 
continuous quantum numbers in the reciprocal space defined modulo
a translation of the reciprocal lattice~$\widetilde{\mathcal{R}}$. To obtain the complete spectrum
and the associated eigenbasis, it is necessary and sufficient for~$\tv{k}$ to run through
an elementary cell that we will choose, for instance, to be the first 
Brillouin zone~$\widetilde{\mathcal{C}}$. Bloch theorem \citep[chap.~8, for instance]{Ashcroft/Mermin76a}
essentially says that the Hilbert space~$\mathscr{H}$ of the states of the system can 
be decomposed in a direct sum of subspaces~$(\mathscr{H}_{\,\tv{k}})_{\tv{k}\in\widetilde{\mathcal{C}}}$ where,
for a given~$\tv{k}$, with
\begin{equation}
   \hatW(\tv{k})\DEF\EXP{-\imat \hat{\tv{r}}\cdot\tv{k}}
\end{equation}
being the unitary translation operator by~$-\hbar\tv{k}$ in the reciprocal space,  
the discrete eigenvalues of
\begin{equation}
  \hat{H}_{\tv{k}}\DEF \hatW(\tv{k})\hat{H}\hatW^*(\tv{k})= H(\hat{\tv{p}}+\hbar\tv{k},\hat{\tv{r}})
\end{equation}
 are precisely labelled by~$\sigma$ and
allow to reconstitute the whole spectrum~$E_\sigma(\tv{k})$. The associated eigenvectors of~$\hat{H}_{\tv{k}}$
are given by the Bloch states
\begin{equation}
  \ket{u_\sigma(\tv{k})}\DEF\hatW(\tv{k})\ket{\phi_\sigma(\tv{k})}
\end{equation}
that is
\begin{subequations}
\begin{equation}
   \hat{H}_{\tv{k}}\ket{u_\sigma(\tv{k})}=E_\sigma(\tv{k})\ket{u_\sigma(\tv{k})}
\end{equation}
and are strictly $\mathcal{R}$-periodic,
\begin{equation}\label{eq:TRu}
  \hat{T}(\mathbf{R})\ket{u_\sigma(\tv{k})}=\ket{u_\sigma(\tv{k})}
\end{equation}
\end{subequations}
in contrast with~\eqref{eq:TRpsi}. The corresponding wavefunctions~$\phi_{\sigma,\tv{k}}(\tv{r})\DEFt\braket{\tv{r}}{\phi_\sigma(\tv{k})}$ and the associated Bloch functions~$u_{\sigma,\tv{k}}(\tv{r})\DEFt\braket{\tv{r}}{u_\sigma(\tv{k})}$
are related by
\begin{equation}\label{eq:phiblochubis}
  \phi_{\sigma,\tv{k}}(\tv{r})=\EXP{\imat \tv{k}\cdot\tv{r}}u_{\sigma,\tv{k}}(\tv{r})
\end{equation}
whereas~\eqref{eq:TRu} reads, for any~$\tv{R}\in\mathcal{R},$
\begin{equation}\label{eq:TRubis}
  u_{\sigma,\tv{k}}(\tv{r}+\tv{R})=u_{\sigma,\tv{k}}(\tv{r})\;.
\end{equation}
Consider the propagator given by~\eqref{eq:Kpsipsi} where the sum is restricted to~$\mathscr{H}_{\,\tv{k}}$,
\begin{align}
  K_{\tv{k}}(\tv{r}_f,t_f,\tv{r}_i,t_i)&=\sum_{\sigma}
 \phi^{}_{\sigma,\tv{k}}(\tv{r}_f)\phi^*_{\sigma,\tv{k}}(\tv{r}_i)\,\EXP{-\frac{\imat}{\hbar}(t_f-t_i)\, E_\sigma(\tv{k})};\\
&=\sum_{\sigma} u^{}_{\sigma,\tv{k}}(\tv{r}_f)u^*_{\sigma,\tv{k}}(\tv{r}_i)\,\EXP{\imat \tv{k}\cdot(\tv{r}_f-\tv{r}_i)-\frac{\imat}{\hbar}(t_f-t_i)\, E_\sigma(\tv{k})}\;.
\end{align} 
Now the~$\widetilde{\mathcal{R}}$-periodicity of the bands~$\tv{k}\mapsto E_\sigma(\tv{k})$
and~$\tv{k}\mapsto \phi^{}_{\sigma,\tv{k}} $ allows to expand~$K_{\tv{k}}$ into \nompropre{Fourier} series
according to:
\begin{equation}\label{eq:Kksommeclasse}
   K_{\tv{k}}(\tv{r}_f,t_f,\tv{r}_i,t_i)=\sum_{\tv{R}\in\mathcal{R}}\EXP{-\imat \tv{k}\cdot\tv{R}}K_{\tv{R}}(\tv{r}_f,t_f,\tv{r}_i,t_i)
\end{equation}
with ($\tilde{v}$ stands for the volume of~$\widetilde{\mathcal{C}}\,$)
\begin{subequations}\label{eq:Kksommeclassebis}
\begin{align}
  \hspace{-1ex}K_{\tv{R}}(\tv{r}_f,t_f,\tv{r}_i,t_i)&=\frac{1}{\tilde{v}}
\int_{\widetilde{\mathcal{C}}} K_{\tv{k}}(\tv{r}_f,t_f,\tv{r}_i,t_i)\,
\EXP{\imat\tv{k}\cdot\tv{R} }\dmat^\Dim\tv{k};\\
&=\frac{1}{\tilde{v}}\int_{\widetilde{\mathcal{C}}}\sum_{\sigma}\hspace{-1ex}
\underbrace{u^{}_{\sigma,\tv{k}}(\tv{r}_f)}_{=u_{\sigma,\tv{k}}(\tv{r}_f+\tv{R})}\hspace{-1ex}u^*_{\sigma,\tv{k}}(\tv{r}_i)
\,\EXP{\imat \tv{k}\cdot(\tv{r}_f-\tv{r}_i+\tv{R})-\frac{\imat}{\hbar}(t_f-t_i) E_\sigma(\tv{k})}
\dmat^\Dim\tv{k};\\
&=\frac{1}{\tilde{v}}
\int_{\widetilde{\mathcal{C}}} \sum_{\sigma}
\phi^{}_{\sigma,\tv{k}}(\tv{r}_f+\tv{R}) \phi^*_{\sigma,\tv{k}}(\tv{r}_i)
\,\EXP{-\frac{\imat}{\hbar}(t_f-t_i) E_\sigma(\tv{k})}
\,\dmat^\Dim\tv{k};\\
&=\barK(\tv{r}_f+\tv{R},t_f,\tv{r}_i,t_i).\label{eq:KbarK}
\end{align}
\end{subequations}
In the antepenultimate expression  we have recognised  the full propagator~$\barK$, \textit{i.e.} built with the complete
spectrum of~$\hat{H}$, for the particle going from~$\bar{q}_i=\tv{r}_i$ to~$\bar{q}_f=\tv{r}_f+\tv{R}$.
In fact, $K_{\tv{R}}(\tv{r}_f,t_f,\tv{r}_i,t_i)$ or~$\barK(\tv{r}_f,t_f,\tv{r}_i,t_i)$ are generally 
not $\mathcal{R}$-periodic, even up to a phase, neither in~$\tv{R}$, neither in~$\tv{r}_f$, neither in~$\tv{r}_i$  but rather, 
for all~$\tv{R}'\in\mathcal{R}$,
from~\eqref{eq:TRubis},
\begin{eqnarray}\label{eq:KripR}
K_{\tv{R}}(\tv{r}_f+\tv{R}',t_f,\tv{r}_i,t_i)=K_{\tv{R}}(\tv{r}_f,t_f,\tv{r}_i-\tv{R}',t_i)=K_{\tv{R}+\tv{R}'}(\tv{r}_f,t_f,\tv{r}_i,t_i)
\end{eqnarray}
whereas~$K_{\tv{k}}$ inherits of the boundary conditions satisfied by~$\phi_{\sigma,\tv{k}}$:
\begin{subequations}
\begin{align}
   K_{\tv{k}}(\tv{r}_f+\tv{R},t_f,\tv{r}_i,t_i)&
=\EXP{\imat\tv{k}\cdot\tv{R}}K_{\tv{k}}(\tv{r}_f,t_f,\tv{r}_i,t_i);\\
K_{\tv{k}}(\tv{r}_f,t_f,\tv{r}_i+\tv{R},t_i)&
=\EXP{-\imat\tv{k}\cdot\tv{R}}K_{\tv{k}}(\tv{r}_f,t_f,\tv{r}_i,t_i).
\end{align}
\end{subequations}
In terms of path integrals,~$\barK(\tv{r}_f+\tv{R},t_f,\tv{r}_i,t_i)$
involves all the paths connecting~$(\tv{r}_i,t_i)$ to
$(\tv{r}_f+\tv{R},t_f)$ and, like~$K_{\tv{R}}(\tv{r}_f,t_f,\tv{r}_i,t_i)$, is defined in the whole
(simply-connected)~$\barQ=\RR^\Dim$. In the equality~\eqref{eq:KbarK},
we recover the identity~\eqref{eq:kK} where~$Q$ is a primary
cell~$\mathcal{C}$ of the crystal, ~$g$ is associated to a spatial
translation by~$\tv{R}$ in the Bravais lattice, and the partial
propagator~$k_{\tv{R}}(\tv{r}_f,t_f,\tv{r}_i,t_i)$ defined to be the
restriction of~$(\tv{r}_f,\tv{r}_i)\mapsto
K_{\tv{R}}(\tv{r}_f,t_f,\tv{r}_i,t_i)$ to~$\mathcal{C}$.  As explained
in the introduction, one can identify an open set of~$Q$ with an open
set of its universal covering space~$\barQ$. This is done naturally
when restricting~$\bar{q}=q=\tv{r}$ to the interior of~$\mathcal{C}$;
yet, as soon as we add~$\tv{R}\neq\tv{0}$ to such an~$\tv{r}$ we get
outside this identification zone.  The configuration space~$Q$ is obtained
by identifying in~$\barQ$ every two points~$(\tv{r},\tv{r}')$ if 
and only if~$\tv{r}'-\tv{r}\in\mathcal{R}$ . Then, $Q$~reduces to
the primary cell~$\mathcal{C}$ with its opposite boundary edges
identified; it is obtained by quotienting~$\barQ=\RR^\Dim$ by the
commutative $\mathcal{R}$-translation group which is then interpreted
as the first homotopy group~$\ZZ^\Dim$ of the~$\Dim$-torus thus
obtained. The relations~\eqref{eq:KripR} when one of
the~$(\tv{r}_f,\tv{r}_i)$ lies on one edge of the boundary
of~$\mathcal{C}$ illustrate what we have generally established, namely
the discontinuity of the~$k$'s.  Taking~$t_f\to t_i$ we have, as an
illustration of~\eqref{eq:enKbar},
\begin{equation}
    \barK(\tv{r}_f+{\tv{R}},t_i,\tv{r}_i,t_i)=\delta(\tv{r}_f+\tv{R}-\tv{r}_i)
\end{equation} 
a well-defined distribution in~$\barQ$ but that becomes problematic when tried to be
restricted to~$\mathcal{C}$, see~\eqref{eq:kneeqdelta},
because it is not $\mathcal{R}$-periodic even up to a phase unlike 
\begin{equation}
   K_{\tv{k}}(\tv{r}_f,t_i,\tv{r}_i,t_i)=\sum_{\tv{R}\in\mathcal{R}}\EXP{-\imat \tv{k}\cdot\tv{R}}\delta(\tv{r}_f+\tv{R}-\tv{r}_i)
\end{equation}
whose restriction to the \emph{interior} of~$\mathcal{C}$ coincides with~$\delta(\tv{r}_f-\tv{r}_i)$
 in agreement 
with~\eqref{eq:elementneutreK}.
Now, when~$(\tv{r}_f,\tv{r}_i)\in\mathcal{C}^2$, and when folding each path~$\bar{\mathscr{C}}$ in~$\barQ=\RR^\Dim$
to a path~$\mathscr{C}$ in~$Q$, the Fourier series~\eqref{eq:Kksommeclasse} 
is exactly an expansion of
the form~\eqref{eq:Kdecomp}. The Bloch angles~$\tv{k}$ label the $\mathrm{U}(1)$-representation 
of~$\pi_1(Q)=\ZZ^\Dim$ 
\begin{equation}
  E(\tv{R})=\EXP{-\imat \tv{k}\cdot\tv{R}}\;.
\end{equation} 
\citet[eqs.~(2.2) \& (2.6)]{Schulman69a} has proposed the decomposition~\eqref{eq:Kksommeclasse}
but did not explicitly interpreted~$K_{\tv{R}}$ beyond of being a simple Fourier coefficient, all the more
that he is sticking to Green functions rather than matrix elements of the evolution operator (without a 
time Heaviside function). However, surprisingly, to my knowledge, in all  texts
 that try to introduce the path-integral formulation in a multi-connected space,
including~\citep[\S\,23.1]{Schulman81a}, this general Bloch framework is abandoned
to exemplify~\eqref{eq:Kdecomp} for the sole free motion on the circle ($\Dim=1$) (in mathematical physics
see \citealp[and its references]{Kocabova/Stovicek08a} that deal exclusively with the Laplace-Beltrami
operator), heavily reinforced with
Poisson summation formulae or Jacobi functions. Again, the choice
of a particular Lagrangian as well as any differential structure,
can only reduce the perspective and mask the fact that we deal with topology.

\section{Homotopy versus homology}\label{sec:homology}

There is another family of groups that provides topological invariants
of the configuration space, namely the homology/cohomology
groups \citep[chap.~2 for instance]{Hatcher02a}. Among those
is the first homology group, traditionaly denoted
by~$H_1(Q)$, which is made of one-dimensional cycles (loops with a
moveable basepoint)
that are not boundaries of a two-dimensional surface included in~$Q$;
two chains being equivalent if they define the boundary of a
two-dimensional surface in~$Q$. A non-unit element of~$H_1(Q)$ is
typically the equivalence class of a chain around a ``hole'' in~$Q$
and therefore both the groups~$\pi_1(Q)$ and~$H_1$ probe the ``holes''
in~$Q$. However, they are generally different since unlike the first
one, $H_1$~is always commutative \citep[\S\,2A]{Hatcher02a}.

With the use of Stokes theorem, one physically associates~$H_1(Q)$
with the magnetic flux through the hole obtained by computing the
circulation of a vector potential~$A$ (a one-form) along a cycle~$c$
in~$H_1$: for a unit electric charge,
\begin{equation}
\Phi=\int_c A\, \dmat x
\end{equation}
and, obviously, the position of a base point chosen to compute the integral along a loop is irrelevant.
This is the kind of topological phase that plays a key role in Dirac's work on magnetic monopole and
in the Ehrenberg-Siday-Aharonov-Bohm effect both mentioned in \S\,\ref{sec:intro}.
As long as we work with factors~$E$ that are in~$U(1)$, the physical properties coming from~\eqref{eq:Kdecomp}
cannot discriminate between~$H_1(Q)$ and~$\pi_1(Q)$: the commutativity of the phases is not able to
reflect the non-commutativity of~$\pi_1(Q)$ and
if we want a finer signature of this non-commutativity we must consider systems
whose~$E$ are unitary matrices of dimension at least~$2$.
One can also understand this requirement in figure~\eqref{fig:2tori}: b)~and c)~are homotopically different (you cannot move the base point)
but homologically identical (when you can move the base point)
since the magnetic flux through them is the same, namely the magnetic flux carried by the left torus only.

Even in the case of anyons, one is unable to say if the topological properties at stakes are the
ones of~$\pi_1$ rather that~$H_1$. Anyons can be interpreted as scalar particles moving in
a two-dimensional surface, each of them carrying an individual
magnetic flux~$\Phi$ perpendicular to the surface. As they classically evolve with an interaction that prevents them from
being at the same place at the same time, the trajectories of each anyon accumulate  an  Ehrenberg-Siday-Aharonov-Bohm phase
while wrapping one around each other in a braid-like structure. Actually, the braid group with~$N$ strands is precisely the~$\pi_1$
of the configuration space of~$N$ anyons and it is not commutative as soon as~$N\geq3$. If~$\mathfrak{b}_n$ stands for
the generator of the braid group where particles~$n$ and~$n+1$ are exchanged, we actually have the Artin-Yang-Baxter relation
\begin{equation}\label{eq:AYB}
  \mathfrak{b}_n\mathfrak{b}_{n+1}\mathfrak{b}_n=\mathfrak{b}_{n+1}\mathfrak{b}_{n}\mathfrak{b}_{n+1}\;.
\end{equation}
For a set of identical anyons,
the phase of each generator
of the braid group~$E(\mathfrak{b}_n)$ is independent of~$n$ for~\eqref{eq:AYB}
to be satisfied and can be taken to be~$\EXP{\imat\Phi/(2\hbar)}$ but,
again, the non-commutativity of the braid group is lost. To recover it, one should not only
work with a model of particles  with an internal degree of freedom such that their propagator
involves not pure topological phases but unitary matrices~$E_n=E(\mathfrak{b}_n)$ but also
accept to deal with~$E_n$ that depends on~$n$ which is hardly sustainable for identical particles.
However, when different species of anyons are present, one may recover some properties
that emerge from the non-Abelian character of their intertwining (see \citealp{Nayak+08a} for a pedagogical
review
of these so called non-Abelian anyons).

Coming back to a configuration space with two holes, each of them being associated with one generator
of~$\pi_1(Q)$ :
we can easily  conceive an experimental set up where this is relevant, with two superconducting tori 
or two Mach-Zehnder interferometers can be used. One can even think of a representation of~$\pi_1$
still keeping its non-commutativity but having a finite number of images provided by~$E$.
The smallest non-commutative finite group is the permutation group of~$3$ elements and the smallest
dimension of a unitary non-commutative linear representation of it is~$3$ (as in the 1D-case, the constraints imposed
by~$\tau^2=1$ for any transposition~$\tau$  that generates the group
force the 2D unitary matrices to be~$\pm1$ and therefore commutative). For instance, up to any global rotation, to represent
two transpositions one can choose
\begin{equation}
  E_1\DEF\begin{pmatrix}-1&0&0\\0&0&-1\\0&-1&0\end{pmatrix};\qquad E_2\DEF\begin{pmatrix}0&0&-1\\0&-1&0\\-1&0&0\end{pmatrix}
  \end{equation}
 as the two non-commutative generators. The third transposition is represented by
\begin{equation}
    E_3\DEF E_2E_1 E_2=E_1E_2 E_1=\begin{pmatrix}0&-1&0\\-1&0&0\\0&0&-1\end{pmatrix}
\end{equation}
and the circular permutations are represented by
\begin{equation}
  E_+\DEF E_1E_2=\begin{pmatrix}0&0&1\\1&0&0\\0&1&0\end{pmatrix};\qquad E_-\DEF E_2E_1=\begin{pmatrix}0&1&0\\0&0&1\\1&0&0\end{pmatrix};
\end{equation}
On a three dimensional Euclidean vectors all the~$E$'s belong to~$SO(3)$;
the~$E_1$, $E_2$, $E_3$ are rotations of angle~$\pi$ around the axis
$\left(\begin{smallmatrix}\;0\\\;1\\\!\!\!-1\end{smallmatrix}\right)$, $\left(\begin{smallmatrix}\!\!\!-1\\\;0\\\;1\end{smallmatrix}\right)$ and~$\left(\begin{smallmatrix}\;1\\\!\!\!-1\\\;0\end{smallmatrix}\right)$ respectively,  and~$E_\pm$
are rotations of angle~$\pm 2\pi/3$ around the axis~$\left(\begin{smallmatrix}1\\1\\1\end{smallmatrix}\right)$. 
\begin{equation}
E_1^2= E_2^2=E_3^2=1;\qquad E_\pm^2=E_\mp;\qquad  E_\pm^3=1\;. 
\end{equation}
Having two independent non commuting generators, say~$\mathfrak{l}_1$ and~$\mathfrak{l}_2$
as in Fig.~\ref{fig:2tori}, $\pi_1(Q)$ is isomophic to the set of the double infinite sequences of integers $(n_1,m_1,\cdots,n_i,m_i\cdots)$
but any  element~$\mathfrak{l}_1^{n_1}\cdot\mathfrak{l}_2^{m_1}\cdots\mathfrak{l}_1^{n_i}\cdot\mathfrak{l}_2^{m_i}\cdots$
 can be unitarily
 represented by one of the six rotations defined above~$\{1,E_1,E_2,E_3,E_+,E_-\}$ if we choose~$E(\mathfrak{l}_1)\DEF E_1$
 and~$E(\mathfrak{l}_2)\DEF E_2$. One may conceive that such rotations may be physically implemented on a spin-1 particle, for instance
 a fictitious spin obtained by working with cold atoms where only a bunch of 3-(sub)levels is relevant to describe their interaction
 with light.

\section*{Acknowedgements}  

It is a pleasure to acknowledge my deep gratitude to Alain Comtet
of the Laboratoire de Physique Th\'eorique et de Mod\`eles Statistiques de l'Universit\'e Paris Saclay 
 for his precious advices on this work; all the more that he somehow  initiated it some decades
ago while guiding my first steps in physics research. Many thanks also to Dominique Delande
for his continuous support and  hospitality at the Laboratoire Kastler-Brossel.

\ifx\undefined\BySame
\newcommand{\BySame}{\leavevmode\rule[.5ex]{3em}{.5pt}\ }
\fi
\ifx\undefined\textsc
\newcommand{\textsc}[1]{{\sc #1}}
\fi
\ifx\undefined\emph
\newcommand{\emph}[1]{{\em #1\/}}
\fi

\end{document}